\documentclass[twocolumn,amsmath,amssymb]{revtex4}
\usepackage{graphicx}
\usepackage{dcolumn}
\usepackage{bm}
\begin{document}

\title{Reentrant fractional quantum Hall states in bilayer graphene:
Controllable, driven phase transitions}
\author{Vadim M. Apalkov}
\affiliation{Department of Physics and Astronomy, Georgia State University,
Atlanta, Georgia 30303, USA}
\author{Tapash Chakraborty$^\ddag$}
\affiliation{Department of Physics and Astronomy,
University of Manitoba, Winnipeg, Canada R3T 2N2}

\date{\today}
\begin{abstract}
Here we report from our theoretical studies that in biased bilayer graphene, 
one can induce phase transitions from an incompressible state to a compressible 
state by tuning the bandgap at a given electron density. Likewise, variation 
of the density with a fixed bandgap results in a transition from the FQHE states
at lower Landau levels to compressible states at intermediate Landau levels 
and finally to FQHE states at higher Landau levels. This intriguing scenario 
of tunable phase transitions in the fractional quantum Hall states is unique 
to bilayer graphene and never before existed in conventional semiconductor systems.
\end{abstract}
\maketitle

The unconventional quantum Hall effect in monolayer graphene, whose experimental 
observation \cite{novoselov_kim} unleashed quite unprecedented interest in this 
system \cite{review}, reflects the unique behavior of massless Dirac fermions in 
a magnetic field \cite{wallace,mcclure}. In bilayer graphene this effect confirms 
the presence of massive chiral quasiparticles \cite{novoselov_bi}. An important 
characteristic of bilayer graphene is that it is a semiconductor with a tunable 
bandgap between the valence and conduction bands \cite{pereira}. This modifies the
Landau level spectrum and influences the role of long-range Coulomb interactions 
\cite{abergel}. Here we report that the fractional quantum Hall effect (FQHE), a 
distinct signature of interacting electrons in the system \cite{fqhe_book,stormer} 
is very sensitive to the interlayer coupling strength and the bias voltage. We 
propose that by tuning the bias voltage, one can induce phase transitions from an 
incompressible state to a compressible state at a given gate voltage. In the same vein,
variation of the gate voltage at a fixed bias potential results in a transition 
from the FQHE states in lower Landau levels to compressible states in intermediate 
Landau levels and finally to FQHE states in higher Landau levels. This interesting 
scenario of tunable phase transitions in the FQH states is unique to bilayer 
graphene and never before existed in conventional semiconductor systems. The FQHE 
in monolayer graphene was in fact, studied theoretically by us \cite{vadim_fqhe} 
and subsequent experiments confirmed the existence of that effect in suspended 
monolayer graphene samples \cite{fqhe_kim}. No such studies have been reported
on bilayer graphene.

We assume that the bilayer graphene consists of two coupled graphene layers with 
the Bernal stacking arrangement. In that case, we are mainly concerned with the 
coupling between the atoms of sublattice A of the lower layer and atoms of 
sublattice B$^{\prime}$ of the upper layer. The single-particle levels in bilayer 
graphene have two-fold spin degeneracy and two-fold valley degeneracy, which can 
be lifted in  many-particle systems at relatively large magnetic fields 
\cite{nakamura}. Considering only one valley (say valley K) and one direction of 
spin, we describe the state of the bilayer system in terms of the four-component 
spinor $(\psi_A, \psi_B, \psi_{B^{\prime}}, \psi_{A^{\prime}})^T$. Here subindices 
A, B and A$^{\prime}$, B$^{\prime}$ correspond to lower and upper layers respectively. 
The strength of inter-layer coupling is described in terms of the inter-layer 
hopping integral, $t$. In a biased bilayer graphene the bias potential is introduced 
as the potential difference, $\Delta U$, between the upper and lower layers. The 
Hamiltonian of the biased bilayer system in a perpendicular magnetic field then takes
the following form
\begin{equation}
{\cal H} =  \left(
\begin{array}{cccc}
    \Delta U/2 & v^{}_F \pi_+ & t & 0   \\
    v^{}_F \pi_- & \Delta U/2 & 0 & 0 \\
    t & 0& -\Delta U/2 & v^{}_F \pi_-   \\
    0 & 0&  v^{}_F \pi_+ & -\Delta U/2
\end{array}
\right),
\label{H1}
\end{equation}
where $\pi_{\pm} = \pi_x \pm  i \pi_y$,  $\vec{\pi} = \vec{p} + e\vec{A}/c$,
$\vec{p}$ is an electron two-dimensional momentum, $\vec{A}$
is the vector potential, and $v^{}_F \approx 10^6$ m/s is the fermi velocity.

In a perpendicular magnetic field the Hamiltonian (\ref{H1}) generates a discrete 
Landau level energy spectrum. The corresponding eigenfunctions can be expressed in
terms of the conventional nonrelativistic Landau functions. The electron states
at sublattices A and A$^{\prime}$ are expressed in terms of the $n$-th `nonrelativistic'
Landau functions, while the electron states at sublattices B and B$^{\prime}$
are described by the $|n-1|$ and $n+1$ Landau functions, respectively. Therefore the 
Landau states in bilayer graphene can be described as a mixture of the $n$, $n+1$, 
and $n-1$ nonrelativistic Landau functions belonging to different sublattices 
\cite{pereira}. This mixture, for a given value of $n$, results in four different 
Landau levels in bilayer graphene. The energies, $\varepsilon $, of the four Landau 
levels corresponding to the index $n$ can be found from the following equation \cite{pereira}
\begin{equation}
\left[\left(\varepsilon + \delta\right)^2 - 2(n+1)\right]
\left[(\varepsilon - \delta )^2 - 2n \right] = (\varepsilon ^2 - \delta ^2 )t^2 ,
\label{level1}
\end{equation}
where $\delta = \Delta U/2$ and all energies are expressed in units of $\hbar v^{}_F/l_B$. 
Here $l_B = (\hbar/eB)^{\frac12}$ is the magnetic length.

It is convenient to introduce special labeling of the Landau levels in bilayer 
graphene. From Eq.~(\ref{level1}), we can see that for each value of $n$, 
$n=0,1,2,\ldots $, there are four solutions, i.e., four Landau levels. Then each 
of these Landau levels can be labeled as $n_{(i)}$, where $i=1,...,4$ is the number 
of the Landau level corresponding to the solution of Eq.~(\ref{level1}) for a given 
value of $n$. 

For a partially occupied Landau level, the properties of the system, e.g., the
ground state and excitations, are completely determined by the inter-electron
interactions, which can be expressed by Haldane's pseudopotentials, $V_m$,
\cite{haldane} which are the energies of two electrons with relative angular
momentum $m$. In a graphene bilayer the Haldane pseudopotentials in a Landau
level with index $n$ and the energy $\varepsilon$ have the form
\begin{equation}
V_m^{(n)} = \int _0^{\infty } \frac{dq}{2\pi} q V(q)
\left[F_{n, \varepsilon } (q) \right]^2 L_m (q^2) e^{-q^2},
\label{Vm}
\end{equation}
where $L_m(x)$ are the Laguerre polynomials, $V(q) = 2\pi e^2/(\kappa l_B q)$
is the Coulomb interaction in the momentum space, $\kappa$ is the
dielectric constant, and $F_{n,\varepsilon} (q)$ are the corresponding form
factors
\begin{eqnarray}
F_{n,\varepsilon}(q) = &  &
d_n^2 \left[
\left(1+ f_n^2  \right) L_n\left(\tfrac{q^2}2\right) +
\frac{2n}{(\varepsilon - \delta )^2} L_{n-1}\left(\tfrac{q^2}2\right)\right.
\nonumber \\
 & & +\left. \frac{2(n+1)}{(\varepsilon + \delta)^2} f_n^2 L_{n+1}\left(
\tfrac{q^2}2\right)\right],
\label{fn}
\end{eqnarray}
where $f_n = [(\varepsilon -\delta )^2 -2n]/[t(\varepsilon - \delta )]$ and
\begin{equation}
d_n = \left[1+ f_n^2 + \frac{2n}{(\varepsilon - \delta)^2 } +
\frac{2(n+1)}{(\varepsilon + \delta)^2} f_n^2 \right]^{-\frac12}.
\end{equation}
The form factors of bilayer graphene [Eq.~(\ref{fn})] are clearly different
from the corresponding form factors of a single layer of graphene. For a
single layer of graphene, the form factor of the $n=0$ Landau level is the 
same as that of conventional non-relativistic electrons 
\cite{vadim_fqhe,goerbig}, $F_{n=0}(q)=L_0\left(q^2/2\right)$. The
form factors of higher Landau levels are determined by the mixture of
$L_n$ and $L_{n+1}$ terms. For bilayer graphene the form factors of the
Landau level with index $n=0$ are the mixtures of the $L_0$ and $L_1$
terms and are different from that in the non-relativistic case. There is of course,
one special Landau level in bilayer graphene with index $n=0$, whose
properties are completely identical to that of the non-relativistic $n=0$
Landau level. Indeed, it is clear from Eq.~(\ref{level1}) that at
$n=0$ there is a Landau level with energy $\epsilon = \delta$. This
energy does not depend on the coupling between the layers, i.e., the
inter-layer hopping integral, $t$. The form factor of this Landau
level is exactly equal to the form factor of a non-relativistic system
of the $n=0$ Landau level, $F_{n=0,\epsilon = \delta} = L_0$. Therefore, 
all many-body properties of a bilayer system in the $n=0$, $\epsilon =
\delta $ Landau level are completely identical to those of a non-relativistic 
conventional system in the $n=0$ Landau level. 

For Landau levels with higher indices the form factor is the mixture
of three different functions, $L_n$, $L_{n-1}$, and $L_{n+1}$. Therefore,
in general, the strength of inter-electron interactions in bilayer
graphene is strongly modified as compared to its value in monolayer
graphene. To address the effects of these modifications on the properties
of the many-electron system in a graphene bilayer we study below the
partially occupied Landau levels with fractional filling factor corresponding 
to the FQHE \cite{fqhe_book}. In what follows, we study the many-electron 
system at various fractional filling factors numerically within the spherical 
geometry \cite{vadim_fqhe,haldane}. The radius of our spehere is
$R=\sqrt{S}l_B$, where $2S$ is the number of magnetic fluxes through the 
sphere in units of the flux quanta. The single-electron states are characterized 
by the angular momentum, $S$, and its $z$ component, $S_z$. For the many-electron 
system the corresponding states are classified by the total angular momentum 
$L$ and its $z$ component, while the energy of the state depends only on $L$ 
\cite{fano}. A given fractional filling of the Landau level is determined by a 
special relation between the number of electrons $N$ and the radius of the sphere
$R$. For example, the $\frac13$-FQHE state is realized at $S = (\frac32)(N-1)$, 
while the $\frac25$-FQHE state corresponds to the relation $S = (\frac54)N-2$.
With the Haldane pseudopotentials [Eq.~(\ref{Vm})] we determine the interaction 
Hamiltonian matrix \cite{fano} and then calculate a few lowest eigenvalues
and eigenvectors of this matrix. The FQHE states are obtained when the
ground state of the system is an incompressible liquid, the energy spectrum 
of which has a finite many-body gap \cite{fqhe_book,stormer}. 

\begin{figure}
\begin{center}\includegraphics[width=8cm]{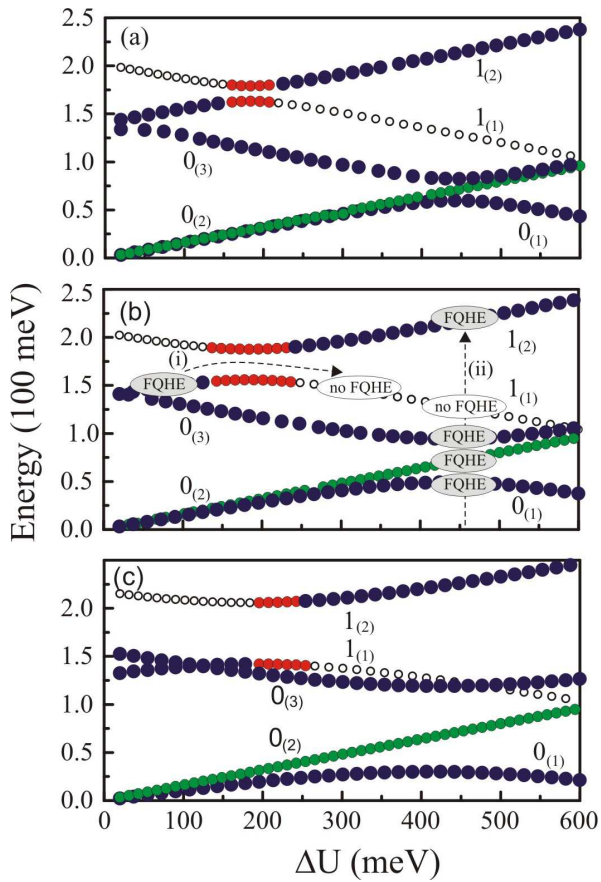}\end{center}
\caption{
A few lowest Landau levels of the conduction band are shown as a
function of the bias potential, $\Delta U$, for different values of
inter-layer coupling: (a) $t=100$ meV (b) $t=200$ meV and (c) $t=400$ 
meV and the magnetic field is 15 Tesla. The numbers next 
to the curves denote the corresponding Landau levels. The Landau levels 
where the FQHE can be observed are drawn as blue and green dots. The 
green dots correspond to the Landau levels where the FQHE states are 
identical to that of a monolayer of graphene or a non-relativistic 
conventional system. The red dots represent Landau levels with weak 
FQHE. The open dots correspond to the Landau levels where the FQHE 
cannot be observed. In (b), the dashed lines labeled by (i) and (ii) 
illustrate two situations: (i) under a constant gate voltage and 
variable bias potential; (ii) under a constant bias potential and 
variable gate voltage. 
}
\label{figone}
\end{figure}

We begin our study with the celebrated $\frac13$-FQHE \cite{stormer} corresponding 
to the filling factor $\nu =\frac13$. In Fig.~\ref{figone} we show the behavior of 
the Landau level spectra as a function of the bias voltage and for different values 
of the inter-layer hopping integral, $t$. Only the Landau levels with positive
energies (corresponding to the conduction band) are shown in the figure. A similar 
behavior is valid for other FQHE filling factors, e.g., for $\nu=\frac25$. 
Figure~\ref{figone} clearly illustrates that the FQHE can be observed in all $n=0$ 
Landau levels with the strongest FQHE being in the first and the
third $n=0$ Landau levels, i.e., $0_{(1)}$ and $0_{(3)}$.
An interesting behavior was also observed for the $n=1$ Landau levels.
For $\Delta U \approx 200$ meV the $n=1$ Landau levels show an {\it
anti-crossing}, which is accompanied by strong changes in the properties
of the FQHE. More specifically, at a small $\Delta U$ the FQHE can
be observed only in the lower $n=1$ Landau level, while for larger $\Delta U$
the FQHE is possible only in higher $n=1$ Landau level. 
These have important implications for possible experimental observations
of this unique behavior (see Fig.~\ref{figone}(b)):


(i) By applying a gate voltage the electron density can be tuned so that
the first three Landau levels in bilayer graphene are completely occupied
and the next Landau level is partially occupied with the FQHE filling factor,
for example, $\nu =\frac13$. According to the notations of Fig.~\ref{figone}, this means
that the $0_{(1)}$, $0_{(2)}$, $0_{(3)}$ Landau levels are fully occupied, 
while the $1_{(1)}$ Landau level has a filling factor $\frac13$. Then, by 
varying $\Delta U$ from a small value, e.g., 100 meV, to a larger value, e.g.,
400 meV, one can observe the disappearance of the FQHE (see line (i) in Fig.~\ref{figone}(b)).

(ii) The bias voltage is kept fixed at a large value, e.g., $\Delta U = 400$ meV.
Then by varying the gate voltage and thus increasing the electron density, one can
observe the disappearance and reappearance of the FQHE at large Landau levels (see
line (ii) in Fig.~\ref{figone}(b)).

The appearance or suppression of the FQHE in different Landau levels is
correlated with the unique behavior of Haldane pseudopotentials in bilayer
graphene. The FQHE is observed in Landau levels with a rapid decrease of the
corresponding pseudopotentials with increasing angular momentum, $m$.
For the first and second $n=0$ Landau levels ($0_{(1)}$ and $0_{(2)}$ levels),
between $m=1$ and $m=3$, the pseudopotential has the fastest decay for the
first level, which results in a higher FQHE gap in this Landau level.
The pseudopotentials in the $0_{(2)}$ Landau level exactly coincide with the
pseudopotentials of the $n=0$ Landau level in a single layer of graphene.
For the $n=1$ Landau levels, the pseudopotentials in one of the $n=1$ levels are 
close to those in the $n=1$ Landau level of the monolayer graphene, resulting in 
a well pronounced FQHE. While in the other $n=1$ Landau level of the bilayer graphene 
the pseudopotentials display a slow decrease with $m$ and no FQHE is expected in 
that Landau level. 


\begin{figure}
\begin{center}\includegraphics[width=9cm]{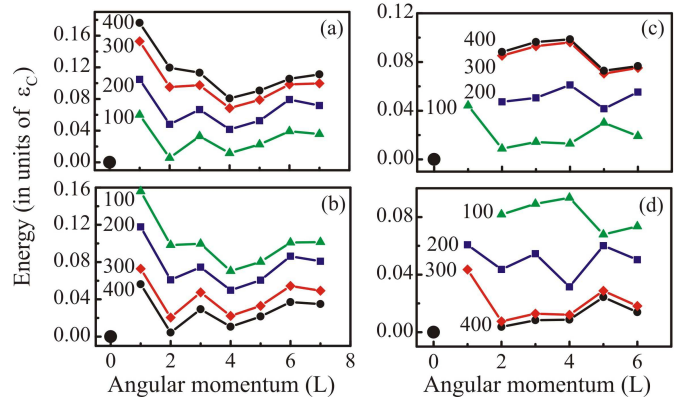}\end{center}
\caption{Low energy excitation spectra of the FQHE states shown for
different values of the bias potential. The numbers next to the
lines are the values of the bias potential in meV.
(a) $\nu =\frac13$-FQHE (eight electrons) system in the Landau level $1_{(2)}$,
(b) $\nu =\frac13$-FQHE (eight electrons) system in the Landau level $1_{(1)}$,
(c) $\nu =\frac25$-FQHE (ten electrons) system in the Landau level $1_{(2)}$,
(d) $\nu =\frac25$-FQHE (ten electrons) system in the Landau level $1_{(1)}$.
The systems are fully spin-polarized. The flux quanta is
$2S = 21$. The magnetic field is 15 T. The ground state is shown as a solid
dot at $L=0$.
}
\label{figfour}
\end{figure}

The collapse of the FQHE gap, corresponding to the appearence of anticrossing
of the $n=1$ Landau levels, is illustrated in Fig.~\ref{figfour}. The FQHE gap 
has a monotonic dependence
on the bias voltage. In the anticrossing region the gap disappears for the lower
$n=1$ Landau level (see Fig.~\ref{figfour}a,c) and reappears for the higher $n=1$ Landau level
(see Fig.~\ref{figfour}b,d). The evolutions of the energy spectra of the incompressible liquid
with different filling factors are similar, which is illustrated in Fig.~\ref{figfour}a,b
(for $\nu =\frac13$) and Fig.~\ref{figfour}c,d (for $\nu =\frac25$). This behavior was
never before observed in the FQHE of conventional two-dimensional electron systems.

\begin{figure}
\begin{center}\includegraphics[width=8.5cm]{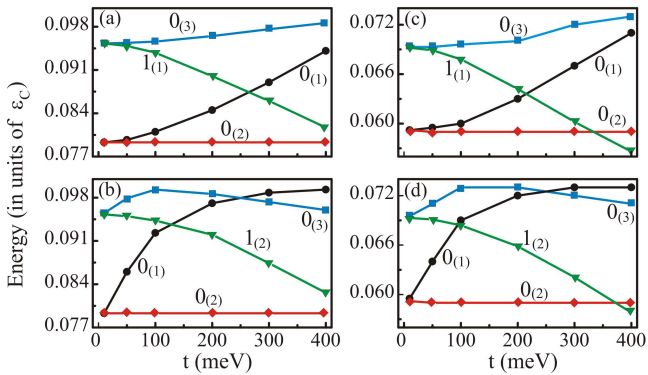}\end{center}
\caption{The FQHE gaps are shown for different Landau levels.
The labels next to the lines correspond to the labeling of Landau
levels shown in Fig.~\ref{figone}.
(a) $\nu =\frac13$-FQHE (eight electron) system at $\Delta U = 10$ meV,
(b) $\nu =\frac13$-FQHE (eight electron) system at $\Delta U = 400$ meV,
(c) $\nu =\frac25$-FQHE (ten electron) system at $\Delta U = 10$ meV,
(d) $\nu =\frac25$-FQHE (ten electron) system at $\Delta U = 400$ meV.
All systems are fully spin polarized. The flux quanta is
$2S = 21$. The magnetic field is 15 Tesla.
}
\label{figfive}
\end{figure}

The strength of the FQHE effect, i.e., the magnitude of the excitation gap,
depends on the parameters of the bilayer graphene, i.e., on the bias voltage
and the inter-layer hopping integral, $t$. In Fig.~\ref{figfive} such dependence is shown
for $\frac13$- and $\frac25$-FQHE in different Landau levels as a function of
the hopping integral, $t$. In accordance with the properties of Haldane
pseudopotentials, the excitation gap of the $0_{(2)}$ Landau levels does not depend
on the bias voltage and on the inter-layer hopping integral. The corresponding
gap remains constant and is equal to gap of the FQHE in a single layer of
graphene in the $n=0$ Landau level. This gap has the smallest value compared
to that of the other Landau levels. For zero hopping integral the two layers
of graphene become decoupled and the bilayer system becomes identical to a
single layer with additional double degeneracy. This property is clearly seen
in Fig.~\ref{figfive}, where at $t=0$ there are only two doubly degenerate FQHE gaps,
corresponding to $n=0$ and $n=1$ single layer Landau levels.

At a small bias voltage ($\Delta U = 10$ meV, see Fig.~\ref{figfive}a,c), the FQHE gaps
have a monotonic dependence on the hopping integral. The values of the gaps
increases for the $n=0$ Landau levels, while for the $n=1$ Landau level the
gap decreases. At large values of the bias voltage, $\Delta U = 400$ meV (see
Fig.~\ref{figfive}b,d), the FQHE gap in the $0_3$ Landau level has a nonmonotonic dependence
on the hopping integral with a maximum around $t=100$ meV. These results also
illustrate that the FQHE in bilayer graphene can be more stable, i.e., the
corresponding FQHE gap is much larger than in the case of FQHE in a single
graphene layer. As an illustration, the FQHE gaps in the  $0_{(1)}$ and 
$0_{(3)}$ Landau levels at finite values of $t$ are larger than the FQHE gaps 
in a single layer of graphene, i.e., at $t=0$.


Our present work suggests an unique opportunity to tune the reentrant FQHE states. 
The transition from the FQHE state to a gapless compressible state is
continuous. The gap decreases monotonically with the bias potential and finally 
collapses. The collapse of the FQHE gap occurs at a non-zero angular momentum, 
$L\approx 2$, resulting in the formation of the ground state with a finite momentum. 
Experimental observation of this transition will provide unique opportunities, 
for the first time, to study the phase transition from an incompressible liquid to 
a possible charge density wave state and then to another incompressible state.

We wish to thank David Abergel for very helpful discussions.
The work has been supported by the Canada Research Chairs Program
and the NSERC Discovery Grant.

\end{document}